\documentclass[12pt]{iopart}
\usepackage{iopams}

\begin{document}
\title{Product Measure Steady States of Generalized Zero Range Processes}
\author{R L Greenblatt\dag\  and J L Lebowitz\dag\ddag}
\address{\dag Department of Physics and Astronomy, Rutgers University, Piscataway NJ 08854, USA}
\address{\ddag Department of Mathematics, Rutgers University, Piscataway NJ 08854, USA}
\eads{\mailto{rafaelgr@physics.rutgers.edu}, \mailto{lebowitz@math.rutgers.edu}}

\begin{abstract}
We establish necessary and sufficient conditions for the existence of factorizable steady states of the Generalized Zero Range Process.  This process allows transitions from a site $i$ to a site $i+q$ involving multiple particles with rates depending on the content of the site $i$, the direction $q$ of movement, and the number of particles moving.  We also show the sufficiency of a similar condition for the continuous time Mass Transport Process, where the mass at each site and the amount transferred in each transition are continuous variables; we conjecture that this is also a necessary condition.
\end{abstract}

%\submitto{\JPA}
\pacs{0250.Ey, 0540.-a}

\section{Introduction}
The classical zero range process (ZRP) is a widely studied lattice model with stochastic time evolution \cite{EH2005}.  To define the process consider a cubic box $\Lambda \subset \mathbb{Z}^d$ with periodic boundary conditions, i.e.\ a $d$-dimensional torus.  At each site $i$ of $\Lambda$ there is a random integer-valued variable $n_i \in \{0,1,\ldots\}$, representing the number of particles at site $i$.  The time evolution is specified by a function $\alpha_q(n_i)$ giving the rate at which a particle from a site $i$ containing $n_i$ particles jumps to the site $i+q$, where $q$ runs over a set of neighbors $E$ (the most common choice being $E=\{\pm e_1, \pm e_2, \ldots\}$).  The name zero range indicates the fact that the jump rate from $i$ to $i+q$ depends only on the number of particles at $i$.

It is easy to see, when the system is finite, that when $\alpha_q(n)+\alpha_{-q}(n) \geq \delta >0$ for all $n > 0$ (we always have $\alpha(0) =0$) and for many choices of $E$ (in particular any that spans $\mathbb{Z}^d$) then all configurations with a given total particle number $N \equiv \sum_{i \in \Lambda} n_i$ are mutually accessible, and hence there is a unique stationary measure $\tilde{P}_\Lambda(\underline{n};N)$ for each $N$.  Normalized superpositions of these measures yield all of the stationary states of this system.  Conversely, given a measure for which there is a nonzero probability of $N$ particles being present in the system one can obtain $\tilde{P}_\Lambda(\underline{n};N)$ by restricting that measure to configurations with $N$ particles.

The ZRP was first introduced in \cite{Spitzer1970}.  It was assumed there and in most subsequent works that the rates $\alpha_q(n)$ are of the form
\begin{equation}
\alpha_q(n)=g_q\alpha(n)
\label{iso}
\end{equation}
with $g_q$ independent of $n$ and $\alpha(n)$ independent of $q$.  In this case the system has the unique steady state given by \cite{Spitzer1970, Andjel1982}
\begin{equation}
\tilde{P}_\Lambda(\underline{n};N)=C_N\delta\left(\sum_{i \in \Lambda}n_i - N\right) \prod_{i \in \Lambda}p(n_i)
\label{pmeasure}
\end{equation}
where 
\begin{equation}
p(n)=\frac{c \lambda ^n}{\prod_{k=1}^n \alpha(k)}
\label{singlesite}
\end{equation}
$C_N$ and $c$ are normalization constants given by
\begin{eqnarray}
c = \left(\sum_{n=0}^\infty \frac{\lambda ^n}{\prod_{k=1}^n \alpha(k)}\right)^{-1}\\
C_N = \left(\sum_{\underline{n} : \sum n_i = N} \prod_{i \in \Lambda}p(n_i)\right)^{-1}
\end{eqnarray}
The unique stationary measure $\tilde{P}_\Lambda(\underline{n};N)$ is thus a restriction to configurations with $N$ particles of the product measure $\prod_{i \in \Lambda} p(n_i)$ with single-site distribution $p(n)$.  

In the limit $\Lambda \rightarrow \mathbb{Z}^d$ with $N/|\Lambda| \rightarrow \rho$ the only stationary extremal measures, i.e.\ the only stationary measures with a decay of correlations, are product measures with $p(n)$ given in Equation (\ref{singlesite}) as the distribution of single site ocupation numbers \cite{Andjel1982}.  These states are parameterized by $\lambda$, which plays the role of the fugacity in an equilibrium system, with different values of $\lambda$ corresponding to different expected particle densities $\rho$, where
\begin{equation}
\rho=\sum_{n=1}^\infty n p(n) = c \sum_{n=1}^\infty \frac{n \lambda^n}{\prod_{k=1}^n \alpha(k)}
\end{equation}

Recently there has been a revival of interest in the ZRP.  For certain choices of $\alpha(n)$, for example when $\alpha(n) \sim b/n^2$ for large $n$, the ZRP on $\mathbb{Z}$ exhibits a transition between a phase where all sites almost certainly contain finite numbers of particles to a `condensed' phase where there is a single site containing an infinite number of particles \cite{Evans2000,GSS2003}.  This condensation has attracted attention as a representative of an interesting class of phase transitions in one dimensional non-equilibrium systems, and has also been applied to models of growing networks \cite{EH2005,KLMST2002}.  

Evans, Majumdar and Zia \cite{EMZ2004} have proposed a generalization of the ZRP, called a Mass Transport Model (MTM).  They considered a one dimensional lattice on which there is a continuous `mass' $m_i \geq 0$ at each site, with a parallel update scheme in which at each time step a random mass $\mu_i$, $0 \leq \mu_i \leq m_i$ moves from each site $i \in \mathbb{Z}$ to the neighboring site $i+1$ with a probability density $\phi(\mu,m_i)$.  This process shares many features with the (totally asymmetric) Zero Range Process, in particular the existence of a condensation transition in certain cases \cite{MEZ2005}.  One very significant difference from the ZRP, however, is that the system has a product measure steady state if and only if $\phi(\mu,m)$ satisfies a certain condition.  This gives rise to a similar condition on the sequential update case (which we will call a Mass Transport Process or MTP), where the rate of transitions from site $i$ to $i+1$ is given by $\alpha(\mu,m)$.  This process has a product measure steady state if and only if there exist functions $g$ and $p$ such that
\begin{equation}
\alpha(\mu,m) = g(\mu)p(m-\mu)/p(m)
\label{mt1d}
\end{equation}

An interesting question, then, is whether similar criteria for the existence of a product measure exist for such processes in higher dimension and with movement in both directions allowed.  This question is already relevant for the ZRP.  A particular case in $d=2$, studied in \cite{vanB, GL}, has
\begin{eqnarray}
\alpha_{\pm 1}(n)=\alpha [ 1 - \delta_{n,0} ] \nonumber \\
\alpha_{ \pm 2}(n)=\alpha^{(2)}(n)=\alpha n \label{2drates}
\end{eqnarray}
i.e.\ a constant rate (independent of $n$) per occupied site for moving in the $\pm x$ direction and a rate proportional to $n$ in the $\pm y$ direction.  A treatment of this system based on fluctuating hydrodynamics and computer simulations (originally conducted on a similar but not quite equivalent system, but which we have reproduced on this system) suggests that this particular system has correlations between occupation numbers at different sites a distance $D$ apart decaying according to a dipole power law $D^{-2}$.  This behavior, which is very different from a product measure steady state or its projection (\ref{pmeasure}), is conjectured to be generic for nonequilibrium stationary states of systems with non-equilibrium particle conserving dynamics in $d \geq 2$.
In the present work we prove rigorously that Equation (\ref{iso}) is a necessary and sufficient condition for the existence of product measure steady states for ZRPs, and as a consequence that the system described by (\ref{2drates}) has no product measure steady states.  This condition in turn is a special case of a condition on a class of systems which we call Generalized Zero Range Processes (GZRP), in which we also allow transitions in which more than one particle moves at a time, although we will assume that the number of particles movimg in a single transition is bounded.  The rate now depends on the number of particles $\nu$ which move in the transition as well as the number of particles $n$ at the site before the transition, and so the rates are given by a function $\alpha_q(\nu,n)$ with some $\nu_{max}$ such that $\alpha_q(n,\nu)=0$ whenever $\nu > \nu_{max}$.  The classical ZRPs discussed above are a special case with $\nu_{max}=1$ and $\alpha_q(1,n)=\alpha_q(n)$.  We prove that a necessary and sufficient condition for the GZRP to have product measure steady states is 
\begin{equation}
\alpha_q(\nu,n)=\frac{g_q(\nu)f(n-\nu)}{f(n)}
\label{condition}
\end{equation}
for some non-negative $g_q(\nu)$ and $f(n)$ with $\sum f(n) < \infty$.  This has a clear similarity to (\ref{mt1d}), and when $\nu_{max}=1$ (\ref{condition}) reduces to (\ref{iso}).  

We will prove Equation (\ref{condition}) in the course of finding a weaker result for the continuous-time Mass Transport Process generalized to dimension $d \geq 1$ and to transitions in all directions.  We show that these systems have product measure steady states when $\alpha_q(\mu,m) = g_q(\mu)p(m-\mu)/p(m)$.  This condition is also necessary under certain conditions (generalizing (\ref{mt1d}) to higher dimension) and we conjecture that this is so in all cases.

\section{Factorizability in the Mass Transport Process}
Let $P_\Lambda(\underline{m},t)$ be the time-dependent probability density of finding the system in a particular configuration $\underline{m}$ with mass $m_i$ at site $i \in \Lambda$, $m_i\in (0,\infty)$.  The master equation describing the evolution of $P_\Lambda(\underline{m},t)$ is
\begin{eqnarray}
\fl \frac{\partial P_\Lambda(\underline{m},t)}{\partial t} = \sum_{i \in \Lambda} \left(-\sum_q \int_0^{m_i} d\mu \alpha_q (\mu,m_i)P(\underline{m},t) \right. \\
\left. + \sum_q \int_0^{m_i}d\mu \alpha_{-q}(\mu,m_{i+q}+\mu)P(\underline{m}^{i,q,\mu},t)\right) \nonumber
\label{master}
\end{eqnarray}
where
\begin{equation}
{m_j^{i,q,\mu}} = 
\left\{\begin{array}{ll}m_j, & j \notin \{i,i+q\} \\m_j-\mu, & j=i \\m_j+\mu, & i=i+q\end{array} \right. 
\end{equation}

A stationary state of the system is a distribution $\tilde{P}_\Lambda(\underline{m})$ such that $\partial P_\Lambda(\underline{m},t)/\partial t = 0$ whenever $P_\Lambda(\underline{m},t)=\tilde{P}_\Lambda(\underline{m})$, or equivalently
\begin{equation}
\fl \sum_{i \in \Lambda} \sum_q \int_0^{m_i} d\mu \alpha_q (\mu,m_i)\tilde{P}_\Lambda(\underline{m}) = \sum_{i \in \Lambda} \sum_q \int_0^{m_i}d\mu \alpha_{-q}(\mu,m_{i+q}+\mu)\tilde{P}_\Lambda(\underline{m}^{i,q,\mu})
\label{stationary}
\end{equation}

We wish to find conditions under which there is a $\tilde{P}_\Lambda$ which is factorizable, that is which takes the form of 
\begin{equation}
\hat{P}_\Lambda(\underline{m})=\prod_{i \in \Lambda} p(m_i)
\label{factor}
\end{equation}

Assuming that $\alpha_q(\mu,m) > 0$ for all $m > 0$ and $0 < \mu \leq m$ for at least one $q$ of each pair of opposite directions, the system in a finite torus $\Lambda$ will have a unique steady state corresponding to each value of the total mass $M = \sum m_i$.  Any linear combination of such states is also a solution of (\ref{stationary}).  Given a factorizable steady state, states of definite total mass can be obtained by projecting $\hat{P}$ onto the set of configurations with a particular value of $M$ in analogy with Equation (\ref{pmeasure}).

Assume that there is a factorizable steady state as in (\ref{factor}).  Let $\bar p (s)$ be the Laplace transform of $p(m)$, and let 
\begin{equation}
\phi_q(\mu,s) = \left[1/\bar p (s)\right]\int_0^\infty dm e^{-sm}\alpha_q(\mu,m+\mu)p(m+\mu)
\label{phidef}
\end{equation} 

Note that, since $\alpha_q(\mu,m)=0$ for $m < \mu$,
\begin{eqnarray}
\fl \int_0^\infty dm e^{-sm}\alpha_q(\mu,m)p(m) \nonumber\\
\lo= e^{-s\mu}\int_0^\infty dm e^{-sm}\alpha_q(\mu,m+\mu)p(m+\mu) = e^{-s\mu}\phi_q(\mu,s)\bar p (s)
\label{lemma1}
\end{eqnarray}

We also have
\begin{eqnarray}
\fl \int_0^\infty dm e^{-sm} \int_0^m d\mu \alpha_q(\mu,m)p(m) &= \int_0^\infty d\mu \int_\mu^\infty dm e^{-sm} \alpha_q(\mu,m)p(m) \nonumber \\
&= \int_0^\infty d\mu e^{-s\mu} \int_0^\infty dm e^{-sm} \alpha_q(\mu,m+\mu)p(m+\mu) \nonumber \\
&= \int_0^\infty d\mu e^{-s\mu} \phi_q(\mu,s)\bar p (s)
\label{lemma2}
\end{eqnarray}

Multiplying both sides of (\ref{stationary}) by $\prod_i e^{-s_i m_i}$ and integrating over all $m_i$, we obtain 
\begin{eqnarray}
\fl\sum_{i \in \Lambda,q \in E}\left(\prod_{j \neq i} \bar p (s_j)\right) \int_0^\infty dm_i \int_0^{m_i}d\mu \alpha_q(\mu,m_i)p(m_i)e^{-s_im_i} \nonumber \\
\lo= \sum_{i \in \Lambda,q \in E} \left(\prod_{j \neq i, i+q} \bar p (s_j)\right) \int_0^\infty dm_i \int_0^\infty dm_{i+q} \int_0^\infty d\mu \nonumber \\
\times \alpha_{-q}(\mu,m_{i+q}+\mu)p(m_i-\mu)p(m_{i+q}+\mu) e^{-s_im_i-s_{i+q}m_{i+q}} \label{intermediate}
\end{eqnarray}

Rewriting (\ref{intermediate}) with the aid of (\ref{lemma1}) and (\ref{lemma2}) and cancelling common factors, we obtain
\begin{equation}
\sum_{i \in \Lambda,q} \int_0^\infty d\mu \phi_q(\mu,s_i) e^{-s_i \mu} = \sum _{i \in \Lambda,q} \int_0^\infty d\mu \phi_{-q}(\mu,s_{i+q})e^{-s_i\mu}
\label{mtcondition}
\end{equation}

Equation (\ref{mtcondition}) will be satisfied if (though not only if)
\begin{equation}
\phi_q(\mu,s)=g_q(\mu)
\label{mtconstant}
\end{equation}

In this case Equation (\ref{phidef}) gives
\begin{equation}
\int_0^\infty dt e^{-sm} \alpha_q(\mu,m+\mu)p(m+\mu) = g_q(\mu) \int_0^\infty dm e^{-sm}p(m)
\end{equation}
which by uniqueness of the Laplace transform gives
\begin{equation}
\alpha_q(\mu,m)=g_q(\mu)\frac{p(m-\mu)}{p(m)} \label{mtsimple}
\end{equation}

Equation (\ref{mtsimple}) is a generalization of the comparable formula for the unidirectional case \cite{EMZ2004}.  In this case and in all other cases where, for each $q \in E$, either $\alpha_q \equiv 0$ or $\alpha_{-q} \equiv 0$ and hence either $\phi_q \equiv 0$ or $\phi_{-q} \equiv 0$, there is in Equation (\ref{mtcondition}) only one term which depends on each pair $m_i,m_{i+q}$, and in order for the equation to be satisfied it must depend on only one of them.  This happens only if (\ref{mtconstant}) holds for that $q$, so in these cases Equation (\ref{mtsimple}) gives the only possible rates for which there is an invariant product measure.

Although in general Equation (\ref{mtconstant}) is not the only way of satisfying Equation (\ref{mtcondition}), solutions of this equation only correspond to realizable dynamics when $p$ and $\alpha_q$ are non-negative and normalizable; the resulting restrictions on $\phi_q$ from Equation (\ref{phidef}) are such that it seems unlikely that there are reasonable (indeed any) rates, other than those in (\ref{mtsimple}), which satisfy all of these conditions.

Dynamics for which the system has a factorizable steady state can be found by beginning with some suitable (positive and normalizable) $p(m)$ and then defining $\alpha_q(\mu,m)$ via (\ref{mtsimple}).  For example let
\begin{equation}
p_c(m) = c e^{-cm} \theta(m)
\end{equation}
\noindent where $\theta$ is the Heaviside step function.  The possible transition rates corresponding to $\tilde P(\underline m ) =\prod p_c(m)$ are
\begin{equation}
\alpha_q(m,\mu) = g_q(\mu) e^{c\mu}\theta(m-\mu) = \tilde{g}_q(\mu) \theta(m-\mu)
\end{equation}
\noindent where $\tilde{g}_q$ are arbitrary non-negative integrable functions, i.e.\ the rates $\alpha_q(\mu,m)$ are independent of $m$ as long as $\mu \leq m$.

\section{Reverse processes}
The existence of product measure states of the Mass Transport Process is related to the nature of the reverse process of the Markov process defined by (\ref{master}).  The reverse of a Markov process defined by transition rates $K(\underline{m} \rightarrow \underline{m}')$ with stationary state $\tilde{P}(\underline{m})$ is defined by rates $K^*(\underline{m}' \rightarrow \underline{m})$ given by
\begin{equation}
K^*(\underline{m}' \rightarrow \underline{m}) = \frac{K(\underline{m} \rightarrow \underline{m}')\tilde{P}(\underline{m})}{\tilde{P}(\underline{m}')}
\label{reverse1}
\end{equation}

Examining (\ref{reverse1}), it is clear that the reverse process of the MTP involves only transitions in which a mass $\mu$ moves from a site $i$ to a site $i+q$, $q \in E$.  Denoting the rates of these transitions by $\alpha^*_{i,q}(\underline{m},\mu)$, and assuming that $\tilde{P}(\underline{m})$ is a product measure with single-site weights $p(m)$, (\ref{reverse1}) becomes
\begin{equation}
\alpha_q(\mu,m_i+\mu)p(m_i+\mu)p(m_{i+q}-\mu)=\alpha_{i+q,-q}^*(\mu,\underline{m})p(m)p(m_{i+q}) \label{reverseDef}
\end{equation}
from which we can see that $\alpha^*_{i,q}$ depends only on $q$, $m_i$, and $m_{i+q}$.  We further note that $\alpha^*_{i,q}$ is independent of $m_{i+q}$, and so defines an MTP, if and only if $\alpha_q$ satisfies Equation (\ref{mtsimple}).  We therefore find that the statement that Equation (\ref{mtsimple}) is a necessary condition for an MTP to have product measure steady states is implied by the statement that the reverse process of any MTP is also an MTP.

\section{Factorizability in Generalized Zero Range Processes}
With mass at each site restricted to an integer particle number $n_i$, we can reproduce the analysis in the previous section up to Equation (\ref{mtcondition}).  Denoting the vector of occupation numbers by $\underline{n}$, and the transition rates by $\alpha_q(\nu,n)$, the stationarity condition corresponding to Equation (\ref{stationary}) is
\begin{equation}
\sum_{i \in \Lambda}\sum_{q \in E}\sum_{\nu=1}^{n_i} \left(-\alpha_q(\nu,n_i)\tilde{P}_\Lambda(\underline{n})+\alpha_{-q}(\nu,n_{i+q}+\nu)\tilde{P}_\Lambda(\underline{n}^{i,q,\nu})\right) =0
\label{zrstationary}
\end{equation}

Suppose $\tilde{P}$ is factorizable,
\begin{equation}
\tilde{P}_\Lambda(\underline{n})=\prod_{i \in \Lambda}p(n_i)
\end{equation}
where $p(n)$ is the probability of having $n$ particles at a given site.  Then define the generating function (discrete Laplace transform)
\begin{equation}
\bar{p}(z)=\sum_{n=0}^\infty z^n p(n)
\end{equation}
and let 
\begin{equation}
\phi_q(\nu,z)=\frac{\sum_{n=0}^\infty z^n \alpha_q(\nu,n+\nu)p(n+\nu)}{\bar{p}(z)}
\label{zrphidef}
\end{equation}
Note that $\phi_q(\nu,z) \geq 0$ for all $\nu,z \geq 0$.
The counterpart of Equation (\ref{mtcondition}) is then
\begin{equation}
\sum_{i \in \Lambda,q \in E} \sum_{\nu=1}^\infty z_i^\nu \phi_q(\nu,z_i) = \sum_{i \in \Lambda,q \in E} \sum_{\nu=1}^\infty z_i^\nu \phi_{\nu,-q}(z_{i+q})
\end{equation}

We now exploit the assumption that transitions occur only for $\nu \leq \nu_{max}$.  Then choosing some $i \in \Lambda$ and $q \in E$ and taking the $k$th derivative of the above expression with respect to $z_i$ and $z_{i+q}$ gives 
\begin{equation}
\sum_{\nu=k}^{\nu_{max}}\frac{\nu!}{(\nu-k)!}z_i^{\nu-k}\phi_q^{(k)}(\nu,z_{i+q})+
\sum_{\nu=k}^{\nu_{max}}\frac{\nu!}{(\nu-k)!}z_{i+q}^{\nu-k}\phi_q^{(k)}(\nu,z_i)=0
\label{dvcondition}
\end{equation}

For $k=\nu_{max}$, we have
\begin{equation}
\phi_q^{(\nu_{max})}(\nu_{max},z_i)+\phi_{-q}^{(\nu_{max})}(\nu_{max},z_{i+q})=0 \label{inducStart}
\end{equation}
For (\ref{inducStart}) to hold for all $z_i$ and $z_{i+q}$, both terms on the left-hand-side must be constant, and thus the functions $\phi_{\pm q} (\nu_{max},\cdot)$ are polynomials of degree $\nu_{max}$; being non-negative they must have non-negative leading terms.  Equation (\ref{inducStart}) states that pairs of leading terms of these polynomials must add up to zero and so each must be zero, and therefore the functions $\phi_q (\nu_{max},\cdot)$ are polynomials of degree at most $\nu_{max}-1$.

Now setting $k=\nu_{max}-1$ we find by the same reasoning that the functions $\phi_q(\nu_{max},\cdot)$ and $\phi_q(\nu_{max},\cdot)$ are polynomials of degree at most $\nu_{max}-2$.  Proceeding in this manner we find
\begin{equation}
\phi_q(\nu,z) = g_q(\nu)
\label{zrcondition}
\end{equation}
as a necessary as well as a sufficient condition.  Referring to the definition of $\phi$, this implies that 
\begin{equation}
\alpha_q(\nu,n)=g_q(\nu) \frac{p(n-\nu)}{p(n)}
\label{gzrpcond}
\end{equation}
is a necessary and sufficient condition for the existence of a product measure.

In the case where $\nu_{max}= 1$ and $\alpha_q(1,n)=\alpha_q(n)$, this condition becomes
\begin{equation}
\alpha_q(n)=c_q\frac{p(n-1)}{p(n)}
\end{equation}
This is what we referred to above as the classical ZRP, with the well-known stationary measure \cite{Spitzer1970, Evans2000} discussed in the introduction.

\section{GZRPs on infinite lattices}
It remains to be established that there are not GZRPs on infinite lattices which have product measure stationary states which do not correspond to stationary states of the corresponding GZRP on a finite periodic lattice.  Let $P(\underline{n})$ be a product measure with single-site distribution $p(n)$ which is stationary for rate functions $\alpha$ on $\mathbb{Z}^d$, and let $\Lambda$ be a finite box in $\mathbb{Z}^d$ which is at least two sites thick in all dimensions.  Denote by $\underline{n}_\Lambda$ the configuration of the system inside of $\Lambda$, and let $P(\underline{n}_\Lambda)$ be the marginal distribution of this configuration.  Then we have
\begin{eqnarray}
\fl \frac{d}{dt} P(\underline{n}_\Lambda) = - \sum_{i \in \Lambda} \sum_{q \in E} \sum_{\nu=1}^{n_i} \alpha_q(\nu,n_i) P(\underline{n}_\Lambda) 
+ \sum_{i \in \Lambda} \sum_{q \in E} \sum_{\nu=1}^{n_{i+q}} \alpha_q(\nu,n_i+\nu)P(\underline{n}_\Lambda^{i,q,\nu}) \nonumber\\
 - \sum_{i \in \partial \Lambda} \sum_{q \in E : i+q \in \Lambda} \sum_{\nu=1}^\infty\sum_{n=\nu}^\infty \alpha_q(\nu,n) P(\underline{n}_\Lambda) p(n) \nonumber\\
 + \sum_{i \in \partial \Lambda} \sum_{q \in E : i+q \in \Lambda} \sum_{n=1}^\infty\sum_{\nu=1}^n \alpha_q(\nu,n) P(\underline{n}^{i,q,\nu}_\Lambda) p(n) \nonumber\\
\lo=0 \label{generator}
\end{eqnarray}
where $\partial \Lambda = \{i \in \mathbb{Z}^d \setminus \Lambda|(\exists q \in E)(i+q \in \Lambda)\}$ and 
\begin{equation}
n^{i,q,\nu}_k = \left\{\begin{array}{ll}
n_k,&k \notin \{i,i+q\} \cap \Lambda \\
n_k+\nu,&k = i \in \Lambda\\
n_k - \nu,&k=i+q \in \Lambda\end{array}\right.
\end{equation}
By repeating the procedure used above with Equation (\ref{generator}) in place of Equation (\ref{zrstationary}), it can easily be seen that $\alpha$ and $p$ must satisfy Equation (\ref{dvcondition}) and so $p$ gives a stationary product measure for the rates $\alpha$ on any finite lattice.

\section{Conclusion}

We have shown that there is a straightforward necessary and sufficient condition, Equation (\ref{condition}), for a generalized Zero Range Process to have a product measure steady state.  For Mass Transport Processes, we have found a condition, Equation (\ref{mtcondition}), for the existence of a product measure steady state, which is considerably more opaque than in the GZRP; it is not clear that this is equivalent to the sufficient condition expressed in Equation (\ref{mtsimple}), the counterpart of the condition we have obtained for GZRPs.  We have, however, presented some reasons to believe that it is.

\ack
The work of RLG was supported by Department of Education grant P200A030156-03.  The work of JLL was support by NSF grant DMR 01-279-26, and by AFOSR grant AF 59620-01-1-0154.  We thank P A Ferrari for a fruitful discussion at the Institut des Hautes \'Etudes Scientifiques in Bures-Sur-Yvette, France, where part of this work was done.  We also thank R K P Zia for useful discussions.

\end{document}